\documentstyle[epsf,aps,multicol,tighten]{revtex}
\draft
\begin{document} 

\title{Nonextensive scaling in a long-range Hamiltonian system}                

\author{Celia Anteneodo}

\address{Centro Brasileiro de Pesquisas F\'{\i}sicas,
         R. Dr. Xavier Sigaud 150, \\
         22290-180, Rio de Janeiro, Brazil \thanks{e-mail: celia@cbpf.br}}

\maketitle

\begin{abstract} 
The nonextensivity of a classical long-range Hamiltonian 
system is discussed. 
The system is the so-called $\alpha$-$XY$ model, a lattice of 
inertial rotators with an adjustable parameter $\alpha$ 
controlling the range of the interactions.  
This model has been explored in detail over the last years. For 
sufficiently long-range interactions, namely $\alpha<d$, 
where $d$ is the lattice dimension, it was shown to be 
nonextensive and to exhibit a second order phase transition. 
However, conclusions in apparent contradiction with the 
findings above have also been drawn.  
This picture reveals the fact that there are aspects of the model 
that remain poorly understood.  
Here we perform a thorough analysis, essaying an explanation 
for the origin of the apparent discrepancies. 

\end{abstract}


\begin{multicols}{2} 
\narrowtext
 
\section{Introduction}

Systems of many particles interacting via long-range forces,
although ubiquitous, are not fully understood 
(see for instance \cite{lrange}). 
Special interest in such  systems has arisen recently in connection  with  
the extension of standard statistical mechanics 
proposed by Tsallis \cite{ct}. 
As a prototype to study the dynamics and thermodynamics of
long-range systems, both in equilibrium and non-equilibrium situations, 
a dynamical model with an adjustable interaction range has 
been introduced \cite{at98}.
The model consists in $N$ interacting rotators moving on parallel planes    
and located on a periodical $d$-dimensional hypercubic 
lattice with  unitary spacing.  
Each rotator is fully described by an angle $0 < \theta_i \le 2\pi$ 
and its conjugate momentum $L_i$. 
The dynamics of the system is ruled by the Hamiltonian

\begin{equation} \label{ham0}
H   =  \frac{1}{2} \sum_{i=1  }^N     L_{i}^{2} +
       \frac{J}{2} \sum_{i=1}^N\sum_{j=1,j\neq i}^N 
       \frac{ 1-\cos(\theta_{i}-\theta_{j}) }{r_{ij}^\alpha}
       \equiv K + V\; ,
\end{equation} 
where the coupling constant is $J\geq 0$ 
(we restrict our study to the ferromagnetic case where interactions
are attractive) and, without loss of generality,
moments of inertia equal to one are chosen for all the particles.
Here $r_{ij}$ measures the minimal distance between rotators
located at the lattice sites $i$ and $j$. 
One can associate to each rotator a ``spin  vector"
${\bf m}_i \;=\; (\cos\theta_i,\sin\theta_i)$, 
which allows to define an order parameter 
$ {\bf m} \;=\; \frac{1}{N} \sum_{i=1}^N {\bf m}_i$. 
The Hamiltonian (\ref{ham0}), describing a classical
inertial $XY$ ferromagnet, is usually referred to as $\alpha$-$XY$ model. 
It includes as particular cases the first-neighbor ($\alpha\to\infty$) and 
the mean-field ($\alpha=0$) models. 
Note that this is an {\em inertial} generalization of 
the well known $XY$ model of the statistical physics of magnetism: the 
time evolution is given by the natural dynamics governed by the
Hamilton equations. 

This  prototype of complex long-range behavior has been thoroughly explored 
in the last few years (see for instance \cite{at98,ta00,cgm00}). 
It has been shown that the model presents {\em nonextensive} behavior for 
$\alpha<d$ \cite{at98}. In that domain of $\alpha$ it displays a second order 
phase transition. This result has been exhibited first by means of
numerical computations for the one-dimensional (1D) case \cite{ta00} and
later through analytical calculations for arbitrary $d$ using a
scaled version of the Hamiltonian $H$ \cite{cgm00}. 
However, a recent work \cite{t03} draws conclusions that are in 
disagreement with the previous findings, claiming that the model is extensive 
for all $\alpha$ and that there is no phase transition. 
This apparent contradiction helps to put into evidence that there are aspects of 
the model that remain obscure. 
The lack of a comparitive study as well as of a discussion on the
origin of the discrepancies motivates the present work. It is the purpose 
of this paper to review and complement previous results to elucidate the question.

In order to do that, we use the following methodology. 
We start by solving the equations of motion associated to Hamiltonian $H$: 

\begin{eqnarray} 
\dot{\theta}_i &=&\frac{\partial H}{\partial L_i}\,=\,L_i, \\ \nonumber
\dot{L}_i &=&-\frac{\partial H}{\partial \theta_i} \,=\,-J\sum_{j\neq i}
\frac{\sin(\theta_i-\theta_j)}{r_{ij}^\alpha}, 
\hspace*{8mm} i=1,\ldots,N.
\label{motioneqs} 
\end{eqnarray} 
Numerical integration is performed by means of a symplectic fourth order 
algorithm\cite{symplectic} using a small time step to warrant energy 
conservation with a relative error smaller than $10^{-5}$.  
Equilibrium properties are analyzed by means of time averages (computed 
after a transient) that allow to mimic mirocanonical averages.
In Ref. \cite{t03}, numerical results for the canonical ensemble were 
obtained through standard Monte Carlo simulations.  
Due to ensemble equivalence \cite{micro}, 
both methods are expected to yield the same macroscopic averages  
at thermal  equilibrium.   
Simulations will be supplemented by analytical considerations.

\section{Equilibrium thermodynamics of the $\alpha-XY$ model}

Along this paper we will consider Hamiltonian (\ref{ham0}) 
although many related works in the literature refer to 
a modified version of this Hamiltonian such that the interactions 
are scaled. Since there is a correspondence between both descriptions, 
we will discuss this point to take profit of all the pertinent 
results in the literature. 
To construct the scaled Hamiltonian, let us call it $\widetilde{H}$,  
the coupling coefficient $J$ in $H$ is substituted by 
$J/\widetilde{N}$, where  $\widetilde{N}$ is  
the upper bound of the potential energy per particle, that  
depends on $N$, $\alpha$, and $d$ according to 
$
\widetilde{N} =
\frac{1}{N}
{\displaystyle \sum_i\sum_{j\neq i } }
\frac{1}{r_{ij}^{\alpha} } $.
In the large $N$ limit one has \cite{ta00,sergio}
\begin{equation}
\widetilde{N} (N,\alpha/d) \sim
\left\{ 
\matrix{ 
  N^{1-\alpha/d} & & 0 \le \alpha < d \cr 
          \ln{N} & & \alpha = d \cr 
\Theta(\alpha/d) & & \alpha > d } \right.  
\label{asym} 
\end{equation} 
with $\Theta$ a function of the ratio $\alpha/d$ only, that goes to 2 
as $\alpha/d$ goes to infinity. 
For $\alpha\leq d$, $\widetilde{N}$  depends strongly on $N$. 
Then the representation given by $\widetilde{H}$ may be considered
artificial, since the microscopic coupling coefficient becomes 
$N$-dependent, that is, becomes fed with macroscopic information.   
Anyway, the thermodynamics and the underlying dynamics of 
$\widetilde{H}$ can be trivially mapped onto those of $H$ by 
transforming energy-like quantities through 
$\widetilde{E} \leftrightarrow E/\widetilde{N}$  and characteristic 
times (as long as moments of inertia remain unitary) 
through $\widetilde{\tau} \leftrightarrow \tau \widetilde{N}^\frac{1}{2}$
\cite{at98}.
The usual preference for the scaled form  $\widetilde{H}$ comes from 
the fact that the thermodynamic limit of $E/N$ is always finite and 
no further scalings of either thermodynamical or 
dynamical quantities are needed. 

It has been analytically shown \cite{cgm00}, through canonical calculations 
performed with Hamiltonian $\widetilde{H}$, that the  thermodynamics of 
systems with $\alpha<d$, at the final thermal equilibrium,
is {\em equivalent} to that of its mean-field version 
(the so-called Hamiltonian Mean Field (HMF) \cite{ar95}). 
Such systems display a second order phase transition,
from a low-energy ferromagnetic state to a high-energy paramagnetic one, 
at a certain critical energy per particle 
$\widetilde\varepsilon_c=\widetilde{U}_c/N=0.75 J$. 
This important result for $\widetilde{H}$ analytically confirms
the previous findings\cite{ta00} for the 1D case 
of the original Hamiltonian $H$, just by taking into account the
simple mapping between $\widetilde{H}$ and $H$. 
Since the equilibrium results of the HMF are universal for $\alpha<d$, 
we will summarize them. 
In terms of the magnetization, Hamiltonian $\widetilde{H}$ 
leads to the following caloric curve 
  
\begin{equation} \label{caloric}
\widetilde{U} \;=\; \frac{{N}}{2\widetilde\beta} \;+\; 
\frac{J N}{2} \bigl[ 1-m^2 \bigr],\hspace*{5mm} 
\mbox{with} \hspace*{5mm} 
m=\frac{I_1(\widetilde\beta Jm)}{I_0(\widetilde\beta Jm)},
\end{equation}  
where $\widetilde\beta\equiv 1/\widetilde{T}$ (being  
$\widetilde{T}=2\langle \widetilde{K}\rangle/N$ the temperature 
and having set the Boltzmann constant $k_B=1$) and $I_n$ are 
the modified Bessel functions of order $n$. 
The consistency equation from which the magnetization is extracted 
can be found for instance through 
canonical calculations \cite{ar95}.  
It has a stable solution $m=0$ for 
$\widetilde\beta J<2$ while, for $\widetilde\beta J>2$, 
the zero magnetization solution becomes unstable and a non-vanishing 
$\widetilde\beta$-dependent stable solution arises. 
From (\ref{caloric}), it is clear that the critical value 
$\widetilde\beta_cJ=2$ corresponds to 
$\widetilde\varepsilon_c=0.75 J$.
Notice in Eq. (\ref{caloric}) that, as $m^2\leq 1$ and 
the inverse temperature $\widetilde\beta$ does not depend on $N$, 
then the large $N$ limit of $E/N$ is always finite. 

We will analyze the size dependence of thermal averages. 
We will focus on the range $0\leq \alpha<1$ of 1D lattices 
governed by Hamiltonian $H$. 
In Fig.~1(a), the average magnetization per particle 
$\langle m \rangle$ is represented as a 
function of the energy per particle $U/N$, for $\alpha=0.5$ and 
different system sizes. Clearly, the energy per particle at which the 
system becomes disordered, i.e, at which the magnetization vanishes up to 
finite size corrections, grows with the system size. 
In Fig.~1(b), the same data are represented as a function of 
$U/N\widetilde{N}$. 
Through this scaling, all data sets tend to the same curve in the 
thermodynamic limit, as it has already been shown previously \cite{ta00}.
In the inset of Fig.~1(b), a plot $\langle m \rangle$ vs. $N$, 
for $U/N\widetilde{N}=1.4$ 
and two values of $\alpha$, illustrate that the magnetization 
in the high energy regime decays with the system size as $1/\sqrt{N}$. 
Additionally, data sets for $N=128$ and different values of 
$\alpha \in [0,1)$ were included in Fig.~1(b) to show that
the curve of magnetization vs. $U/N\widetilde{N}$ is
the same for any $\alpha$-$XY$ system with $0\leq\alpha<1$,
up to corrections of order $1/\sqrt{N}$. In particular,
the universal curve coincides with the one for the HMF model
($\alpha=0$), given by Eq. (\ref{caloric}),
once taken into account the mapping
$\widetilde{E}\leftrightarrow E/\widetilde{N}$.
Everything in agreement with the analytical results of Ref. \cite{cgm00}. 
However, concerning the phase transition, there is a risk to fall 
into an endless rhetorical discussion. Strictly speaking,  
there is no ferromagnetic transition, because the critical energy per 
particle $U_c/N=0.75 J \widetilde{N}$ is divergent, as 
asserted in \cite{t03}.   
Nevertheless, the limit $N\to\infty$, despite being an idealized situation, 
must reflect the behavior of finite but large systems in order to be meaningful. 
Ultimately, we are interested in finite-size systems, as real systems are. 
For finite-size $\alpha$-$XY$ systems, with $\alpha<1$, as those that were 
simulated in this work, one can distinguish two regimes: One, at low energies,
where the system is ordered with a magnetization significantly
different from zero and independent from the system size, and another,
a disordered one, with magnetization of order $1/\sqrt{N}$.
A good, representative, thermodynamic limit, reflecting this situation,
can be defined by means of an appropriate scaling, the one allowing
data collapse.
In that case, it results a finite critical energy, $U_c/N\widetilde{N}=0.75J$, 
which plays the same role as the critical energy per particle in 
an extensive system.

\begin{figure}[hb]

\setlength{\unitlength}{1mm}
\begin{picture}(120,120)(0,0)
\put(-12,-10){\epsfxsize=10cm\epsfbox{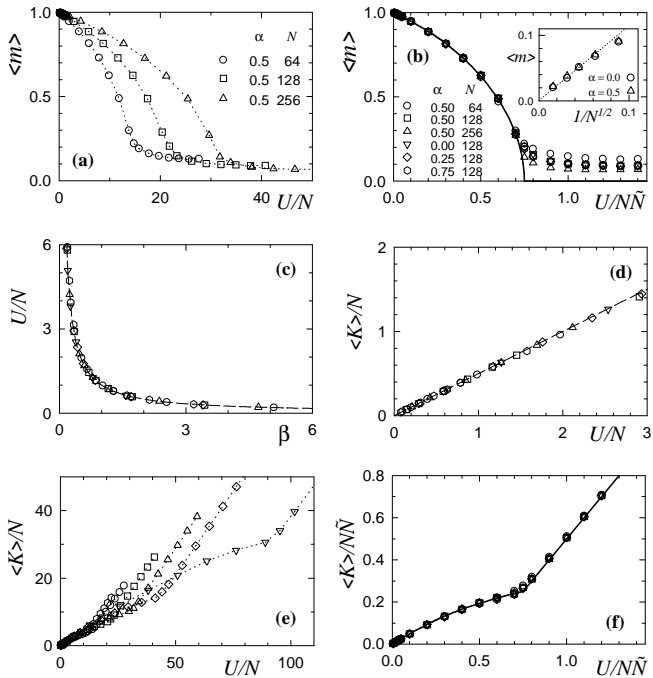}}
\end{picture}
\caption{\protect 
(a)-(b) Average magnetization per particle $\langle m \rangle$ 
as a function of the total energy $U$ for different sets of parameters 
$(\alpha,N)$. 
The inset in (b)  presents the magnetization as a function of $1/\sqrt{N}$ for 
$U/N\widetilde{N}=1.4$. 
(c)-(f) Caloric curves. 
In all cases full lines correspond to the mean-field analytical result 
given by Eq. (\ref{caloric}) and, dotted lines are 
guides to the eyes. Dashed lines correspond to $U/N=1/\beta$ in (c) and 
to a  straight line of slope 1/2 in (d). 
Symbols (defined as in (b)) correspond to averages 
over 10 samples, computed over a time interval of order $10^3$, 
after a transient ($t\approx10^3$) has elapsed and starting from 
"water-bag" [10] initial conditions.  We have set $J=1$. 
The lattice dimension is $d=1$.  } 
\label{fig1} 
\end{figure} 

Let us analyze the dependence of the mean kinetic energy on the total energy 
for different system sizes and different values of $\alpha$. 
In Fig.~1(c) we represent the data as done in Ref. \cite{t03},
that is $U/N$ vs. $\beta$, where $1/\beta=T=2\langle K\rangle/N$. 
In agreement with Ref. \cite{t03}, perfect data collapse occurs for total 
energies scaled with the system size $N$, as can also be appreciated in 
the alternative representation of the same data exhibited in Fig.~1(d). 
However, these plots are restricted to {\em very low energies}. 
If one extends the range of energies plotted (Fig.~1(e)), 
it becomes clear that data collapse does not hold 
any more through the $N$-scaling. 
Whereas, as before, it is the $N\widetilde{N}$-scaling the one which 
leads to data collapse in the  {\em full energy range} (Fig.~1(f)). 
As an aside comment, note that, because the relation
$U\simeq 2\langle K \rangle$ holds at low energies, 
data collapse would occur in that regime for any  arbitrary scaling  
by $N^\gamma$, with $\gamma\in\Re$. 
In particular, this is true for $\gamma=1$, as plotted in Fig. 1(d) 
(hence Fig. 1(e) at low energies) and for $\gamma=2-\alpha$, 
as in Fig. 1(f) at low energies. 

One can understand what is going on as follows. 
For very low energies, the dynamics is dominated by the 
quadratic terms of the potential. 
Thus, the system can be seen as a set of almost uncoupled harmonic 
oscillators (normal modes).
One can also think of particles in a mean-field, a description 
that is exact in the infinite-range case. 
The particles effectively interact not through 
the full mean-field ${\bf m}$ but only through its fluctuations. 
If the mean-field were constant it would play the role 
of an external field and there would be no interactions. 
At low energies, where ${\bf m}$ is almost constant, the residual
or effective interaction, that is the component coming from the 
fluctuations of ${\bf m}$, is small. This is consistent with the 
normal modes view, where interactions are very weak too. 
Therefore, in the limit of very low energies 
(as well as in the limit of very high energies) 
the system becomes non-interacting (hence, integrable). 
While at high energies, i.e., above the critical value, one has almost 
non-interacting rotators;  at low energies, i.e., close to the ground 
state, one has almost non-interacting normal modes. 
Then, at low energies, from the virial theorem, 
the result 
$ \langle K \rangle \simeq \langle V \rangle$  arises trivially. 
The consequent relation $U\simeq 2\langle K \rangle=N/\beta$ 
indicates that the energy is extensive. 
A natural result since the interaction terms are 
not strong, contrarily to what was asserted in \cite{t03}. 
However, as the energy increases and anharmonicities grow, 
the correct scaling choice is no more  
that of an extensive system, as becomes evident in Fig. 1(e). 
Data collapse is actually obtained through the scaling by
$N\widetilde N$, as shown in Figs. 1(b) and 1(f). 
Moreover, this  data collapse is expected to be universal
for any $\alpha\in [0,d)$ \cite{cgm00}.
Hence, at criticality, we have the nonextensive behavior 
$ U_c\propto J N\widetilde{N}$ 
and also $1/\beta_c = T_c \propto J \widetilde{N}$. 

Let us review the whole picture from the viewpoint of canonical
ensemble calculations.
We will consider the case $\alpha=0$, but although tricky, 
a generalization to arbitrary $\alpha \in [0,d)$ could  
be analytically performed \cite{cgm00}. 
The partition function of Hamiltonian (\ref{ham0}) when $\alpha=0$ 
is given by the following integral over phase space
$ Z=\int \prod_{j=1}^{N} dI_j d\theta_j \exp(-\beta H) =Z_K Z_V$, 
which factorizes into the kinetic and potential contributions

\begin{eqnarray} 
Z_K &=& \biggl( \frac{2\pi}{\beta}\biggr) ^{N/2} \hspace*{5mm} \mbox{and}\\ \nonumber
Z_V &=& {\rm e}^\frac{-\beta J N^2}{2} \frac{(2\pi)^N}{\beta J}
\int_0^\infty dy\, {\rm e}^{-NG(y)},
\end{eqnarray} 
where $Z_V$ has already been transformed by means of the 
Hubbard-Stratonovich trick and 
$G(y)=-\frac{1}{N}\ln y + \frac{y^2}{2\beta JN}-\ln I_0(y)$. 
The derivation is the same followed in \cite{ar95} for the scaled 
Hamiltonian $\widetilde H$, apart from an $N$-scaling  that does 
not affect the procedure. 
The integration can be performed by means of the Gaussian approximation 
around the point $y_o$ verifying 
$G'(y_o)=0$, that is, $y_o \simeq\beta J N I_1(y_o)/I_0(y_o)$ and $G''(y_o)>0$.  
In our case, the total energy results

\begin{equation} \label{totalU}
U\;=\; -\frac{\partial \ln Z}{\partial \beta} \;=\;
\frac{N}{2\beta} +\frac{JN^2}{2}(1-m^2),
\end{equation}  
with

$$
m=\frac{I_1(\beta JNm)}{I_0(\beta JNm)},
$$
in correspondence with Eq. (\ref{caloric}). 
For large $\beta J N$, from the consistency equation,
one has $m^2\simeq 1-\frac{1}{\beta J N}$ for the stable solution,
an approximation that is equivalent to considering
$y_o \simeq \beta J N -1/2$,
as done in Ref. \cite{t03}. In fact, substitution of the above approximate 
expression for $m^2$ in (\ref{totalU}), gives $U\simeq N/\beta$. 
Again one obtains that at low temperatures the energy is extensive. 
However, the approximation above is no longer valid as $\beta$ decreases. 
In this case, long-range couplings become effective and the nontrivial 
nonextensive behavior comes out. 
Then, the energy no longer scales with $N$ and one has to consider the 
more general Eq. (\ref{totalU}). 
An analysis as that performed in Ref. \cite{t03}, restricted to the very
low temperature regime, misses most of the rich physics 
of the long-range interacting rotators. 
Of course, this discussion is meaningful as soon as $N$ is not excessively large.
Recall that $1/\beta_c \sim J\widetilde{N}$ for generic $\alpha$,
hence $1/\beta_c \sim JN$  for $\alpha=0$. 
Then $N$ has to be large enough so that the thermodynamic limit is a 
reasonable approximation but not so large as to drive the temperature scale 
out of a realistic range.

\section{Final remarks}

A double sum as in (\ref{ham0}) indicates that,  
for interaction ranges $0 \le \alpha/d < 1$, the total energy $U$ 
may grow as $N^\gamma$, with $\gamma>1$, 
as occurs in the regimes of the 
$\alpha-XY$ model where long-range couplings become relevant 
(see also \cite{kac}). 
Therefore, the large $N$ limit of $U/N$ is not well defined, 
in fact, the energy per particle diverges when $N\to\infty$. 
In that case, the energy is a {\em  nonextensive} quantity \cite{touchette}. 
Many systems in nature also display such kind of behavior, as  
illustrated by Thirring in the context of a discussion on
the stability of matter \cite{thirring}. 
In those cases, it is sometimes said that the thermodynamic limit
does not exist.
However, a proper thermodynamic limit can be effectively achieved by introducing 
a suitable $N$-dependent factor $N^*\sim N^{\gamma-1}$ 
such that the large $N$ limit of $U/N N^*$ results 
{\em well defined} \cite{ct95}. Concerning criticality, 
for the $\alpha-XY$ model, in the thermodynamic limit,  
there is no phase transition in the sense 
that a transition does never occur at a finite energy per particle.   
However for finite-size $\alpha$-$XY$ systems, 
with $\alpha \in [0,d)$, one can distinguish two regimes:
An ordered one at low energies and a disordered one above
a ``critical" energy that increases nonextensively with
the system size (see Fig. 1(a)).
Then, a different limit appears to be the relevant one. 
Indeed, application of an appropriate regularization 
procedure, namely, further scaling by $N^*=\widetilde N$,  
allows to display a transition. 
By means of that scaling, a finite critical energy,
$U_c/N\widetilde{N}=0.75J$, can be defined. 
In this way, a thermodynamic limit, representative  of the behavior 
observed for large $N$ (although not exceedingly), which is  not limited to the 
low energy clustered regime, is obtained.

\section*{Acknowledgments:} 
I am  grateful to S. Ruffo, F. A. Tamarit, C. Tsallis and R.O. Vallejos 
for enlightening discussions. 
I acknowledge financial support from Brazilian agency FAPERJ. 
%


\end{multicols}

\end{document}